\documentclass[conference]{IEEEtran}
\usepackage{epsfig}
\usepackage{acronym}
\usepackage{amssymb}
\usepackage{amsmath}
\usepackage{amsxtra}
\usepackage{booktabs}
\usepackage{multirow}
\usepackage{url}
\usepackage{setspace}
\usepackage{algorithm}
\usepackage{listings}
\usepackage{multirow}
\usepackage[table,xcdraw]{xcolor}
\usepackage{soul}
\usepackage{adjustbox}

\usepackage[utf8]{inputenc}
\usepackage [english]{babel}
\usepackage [autostyle, english = american]{csquotes}

\definecolor{mGreen}{rgb}{0,0.6,0}
\definecolor{mGray}{rgb}{0.5,0.5,0.5}
\definecolor{mPurple}{rgb}{0.58,0,0.82}
\definecolor{backgroundColour}{rgb}{0.95,0.95,0.92}

\lstdefinestyle{customc}{ 
  belowcaptionskip=1\baselineskip,
  breaklines=true,
  xleftmargin=\parindent,
  language=C,
  showstringspaces=false,
  basicstyle=\footnotesize\ttfamily,
  tabsize=2,
  backgroundcolor=\color{backgroundColour},   
  commentstyle=\color{mGreen},
  keywordstyle=\color{magenta},
  numberstyle=\tiny\color{mGray},
  stringstyle=\color{mPurple}
}

\newcommand{\shrink}[1]{{\vspace{-10pt}}}
\newcommand{\comment}[1]{{}}
\newcommand{\forjournal}[1]{{}}
\newcommand{\blind}[1]{{}}
\newcommand{\tout}[1]{{}}

\newcommand{\omitsection}[1]{{\em Section omitted for blind review process.}}
\newcommand{\omitname}[1]{{\em NameOmitted}}
\newcommand{\etal}{{\em et. al.}}

\begin{document}

\title{With ChatGPT, do we have to rewrite our learning objectives - CASE study in Cybersecurity}

\author{
\IEEEauthorblockN{Peter Jamieson}
\IEEEauthorblockA{Department of ECE \\
Miami University, Ohio\\
Email: jamiespa@miamioh.edu}
\and
\IEEEauthorblockN{Suman Bhunia, Dhananjai M. Rao}
\IEEEauthorblockA{Department of CSE \\
Miami University, Ohio\\
Email: \{bhunias, raodm\}@miamioh.edu}
}
\maketitle


\begin{abstract}
With the emergence of Artificial Intelligent chatbot tools such as ChatGPT and code writing AI tools such as GitHub Copilot, educators need to question what and how we should teach our courses and curricula in the future.  In reality, automated tools may result in certain academic fields being deeply reduced in the number of employable people.  In this work, we make a case study of cybersecurity undergrad education by using the lens of ``Understanding by Design'' (UbD).  First, we provide a broad understanding of learning objectives (LOs) in cybersecurity from a computer science perspective.  Next, we dig a little deeper into a curriculum with an undergraduate emphasis on cybersecurity and examine the major courses and their LOs for our cybersecurity program at Miami University.  With these details, we perform a thought experiment on how attainable the LOs are with the above-described tools, asking the key question ``what needs to be enduring concepts?'' learned in this process.  If an LO becomes something that the existence of automation tools might be able to do, we then ask ``what level is attainable for the LO that is not a simple query to the tools?''.  With this exercise, we hope to establish an example of how to prompt ChatGPT to accelerate students in their achievements of LOs given the existence of these new AI tools, and our goal is to push all of us to leverage and teach these tools as powerful allies in our quest to improve human existence and knowledge.

Our general hypothesis is that the LOs of early first- and second-year courses will need to remain as they are in the revised Bloom's Taxonomy verbiage, but the rate at which we can go deeper in the LO can be aided with ChatGPT tools.  Additionally, in our use of UbD to create courses, we believe the exercise will need to be done at all levels of the undergraduate curriculum such that from the LOs and the respective levels we can then devise valuable activities and assessments that will help students develop into valuable thinkers as they navigate their future careers.  We will argue that similar to mathematics where the 3 steps in mathematics (1. Devise 2. Compute 3. Evaluate Correctness of Result) that just like computers have severely reduced the needs of step 2, the same model can be thought of for computing education where ideas and skills such as syntax and algorithmic implementation might be shifted more to the AI-based assistive tools, and the learner/user needs to be able to create the design (analogous to step 1 in mathematics) and evaluate (analogous to step 2 in mathematics) the results produced by the tools.
\end{abstract}

\section{Introduction}

Technological advancements can have an influence on how we approach education, and the emergence of Artificially Intelligent (AI) chatbot tools such as ChatGPT will result in the next wave of debates and questions on how the existence of these tools will change our educational approaches.  ChatGPT is a natural language chatbot tool (where a chatbot is an autonomous robot that will respond to user queries) that has been trained on large language models such that the tool can respond to questions and generate user-requested solutions.  For example, you can ask ChatGPT to write a story in the voice of your favorite author and define a simple plot and characters.  ChatGPT will create that story, and in many cases, the story will be reasonably written.  Additionally, ChatGPT and other tools such as Github's Copilot provide users with the ability to help them design and code in a variety of programming and design languages to build various user-requested functionality.  For example, we have used ChatGPT to design Verilog HDL modules of a Radix-2 SRT divider with reasonable success (noting that the tool is not perfect and expertise is needed for a well-formed design).

These later capabilities have caused those of us who sit in the computing education space into debates about how the existence of these tools changes what and how we teach.  In this work, we propose to approach this debate from the perspective of student learning outcomes (LOs) as defined using verbiage from Bloom's Taxonomy \cite{anderson2001taxonomy}.  With LOs for a particular curriculum (our example case will focus on cybersecurity), we can pose the simple question of how these LOs might change, and what additional LOs should be added to the curriculum.  To aid in this case study, we also will guide our discussion with the help of the useful concept filter described by Wiggins and McTyghe in ``Understanding by Design'' to classify the importance of LOs.  By approaching a curriculum and the courses it comprises with these perspectives, our goal is to demonstrate how we might approach integrating AI tools into our curricula and education.  

For this case study, we use existing cybersecurity organizational schemes to extract the ``enduring understanding'' concepts.  This organizational scheme will then look at some of the LOs asking if the LO can be ``accelerated'' given the available AI tools.  Next, we will look at our existing cybersecurity curricula with courses and look at each LO mapping and where it sits in Bloom's Taxonomy.  This will allow us to see what are the LOs for cybersecurity education and ask the question, ``Should they be leveled up?''  We will provide ChatGPT example usages on how students can be taught to use the tool to deepen their understanding of a knowledge unit.  These examples show that much of what we do as teachers may be offset to ChatGPT given that they learn to use the tool.

\section{Background and Discussion}
\label{sec:background}

Our approach in this paper is to use the lenses of Bloom's Taxonomy \cite{anderson2001taxonomy} and Understanding by Design's \cite{wiggins1998understanding} filter of concepts to understand the Learning Outcomes (LOs) in a particular curriculum and its respective courses.  We have used a similar approach in our previous work \cite{jamiesonGameLense16} to help educators evaluate if they should spend time creating a game to teach a concept. 

Prescribing the approach that student LOs should be written using Bloom's Taxonomy verbs allows us to rank an LO using Bloom's Taxonomy.  This then allows us to question if the existence of an AI tool, which for the remainder of this paper we assume is ChatGPT as of April 2023, changes either the need for the LO or can we accelerate a student learning of the LO and raise the Bloom's Taxonomy verb-level.  Note, this approach is simplistic, and first, we need to have a deeper understanding of Bloom's Taxonomy and a little knowledge of how curricula are created.  

\subsection{Bloom's Taxonomy and Learning Outcomes}

We define learning in this work, using ideas from \cite{zull02brain}, \cite{kolb84learn}: ``Learning is the stabilization of neural networks based on a desired response''.  This definition allows us to relate the idea of learning things to the necessity of trying and doing things, since if the neurons are not firing (by thinking and doing) then your neurons are not wiring (Donald Hebb in 1949 stated, ``neurons that fire together wire together'').  

\begin{figure}
\centering
\center
\includegraphics[width=\linewidth]{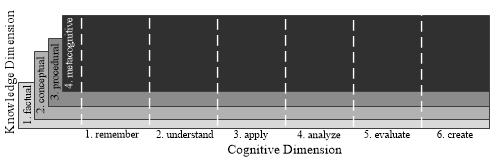}
\caption{The two dimensions of Bloom's taxonomy \cite{krathwohl2002revision}}
\label{bloom}
\end{figure}

A \emph{Learning Outcomes} (LO) is an educational goal for a learner such that they will be able to cognitively perform the objective once they have learned it.  Typically, an LO is described by a cognitive process taken from Bloom's Taxonomy \cite{anderson2001taxonomy} and applying the process to some content as related to the field of study.  Bloom's taxonomy provides a hierarchy of cognitive processes as defined from ``lower-order thinking skills'', such as recall and classification, to ``higher-order thinking skills'' such as creating or planning.  The ``higher-order'' thinking skills tend to be what society wants of our workers, but these higher-order skills are much more challenging to develop and take significantly more time than the lower-order thinking skills to teach and learn because of the profession nature of learning.  Figure \ref{bloom} shows two dimensions of Bloom's taxonomy in terms of cognitive and knowledge dimensions \cite{krathwohl2002revision}.

This distinction between ``higher'' and ``lower'' can, however, confuses educators to somehow evaluate the quality of an LO based on where it sits in the taxonomy.  As illuminated by Lang \cite{lang2021small} professors can fall into the trap of simply wanting to teach ``higher-order thinking skills'' and assume that the lower skills can simply be looked up using technology such as search engines and the like (including tools like ChatGPT).  The reality is to achieve skills such as ``create'' and ``plan'' in Bloom's taxonomy requires that a human can keep and think about ideas and concepts within a field of study that requires the ``remember'' and ``understanding''.  In this work, we are asking this question in terms of accelerating an LO to a higher Bloom's level, but we are aware of this pitfall, and will hopefully provide insight into how the new technology modifies the space.

A university-level course in a particular subject matter will include a number of LOs.  Wiggins and McTighe's book ``Understanding by Design'' \cite{wiggins1998understanding} examines course design by using LOs as the starting point for course design, and then designing a course in reverse from these starting LO goals.  

\subsection{Enduring Concepts and Learning Outcomes}

The LOs need to be defined in a sentence where a verb is taken from Bloom's taxonomy \cite{anderson2001taxonomy} and the object of the sentence describes some knowledge or skill.  For example, a sample LO might be, ``A student should be able to \emph{construct} a program that will then generate a desired output with a proper sequence of program statements''.  This objective is \emph{apply} in Bloom's taxonomy cognitive process dimension \cite{anderson2001taxonomy} and is \emph{procedural} in the knowledge dimension \cite{krathwohl2002revision}.  From the taxonomy, this learning objective would be considered a ``middle-order thinking skill''. 

To further determine how important the LO is, we use Wiggins and McTighe's filter \cite{wiggins1998understanding}:  
\begin{enumerate}
\item{worth being familiar with}
\item{important to know and do}
\item{enduring understanding}
\end{enumerate}
where the importance of a LO is more important to a student's future as the above number increases above.  This filter allows a teacher to consider their LO and qualify how important it is.  The reason this is useful is that just because an objective ranks as a ``low-order thinking skill'' if it is an enduring concept for that student in the respective field, then it is important to emphasize and should be achieved by the student.  Similarly, depending on the time in the curriculum of a field, different ideas will need to be learned to holistically progress in a field.

\subsection{Curriculum Design and Cybersecurity Curricula}

Curriculum design is not a simple process by any means, and for most educators, we can understand this by realizing that a single course is hard to design, and a curriculum is a series of courses; hence it is of greater complexity.  For this reason, we will leave much of the previous research on curriculum design with the exception of noting Sheenan's work \cite{sheehan1986curriculum} that looked at the two conceptual models of the curriculum being either a product model (resulting in a student objective ending) versus the process model (resulting in a student with capabilities).  We argue that the majority of our technical curricula are mostly process models.  

For cybersecurity curricula, a recent paper by Mouheb \etal{}~\cite{mouheb2019cybersecurity} broadly reviews the range of attempts to design cybersecurity curricula.  This broad range of designs sits both in and outside of cybersecurity curricula for computer science.  For the sake of this work, we will focus on undergraduate cybersecurity in computer science to limit the breadth of this space.  Of interest, in Mouheb's work is the three main cybersecurity activities of Prepare, Defend, and Act \cite{rowe2011role} where these activities can be grouped into three distinct phases {\emph{Prevention}}, {\emph{Detection}}, and {\emph{Response}} \cite{riaz2016systematically}.  

Harrison \etal{} \cite{harris2015using} have looked at Bloom's approach to the design of an information technology curriculum as mapped to Bloom's taxonomy.  Their approach to designing the curriculum provides a nice visual representation of how we could approach asking our technological change questions about the existence of ChatGPT.

\section{Learning Outcomes and their link to a Cybersecurity Education}
\label{sec:enduring-concept-inventory}

Using the classification of {\emph{{\b{P}}revention}}, {\emph{{\b{D}}etection}}, and {\emph{{\b{R}}esponse}} we will start by looking at what LOs we believe are part of a modern cybersecurity computer science curriculum.  Here, we note the work on the Cybersecurity Concept Inventory (CCI) which is currently being validated in work by Poulsen \etal{} \cite{poulsen2021psychometric} and by Herman \etal{} \cite{herman2023psychometric}.  Their concept approach identifies concepts that can be validated, and with the validation of this work, an important next step would be having learners progress through our curriculum (ChatGPT enhanced and unenhanced) and evaluate our learners.  

\begin{table}
    \scriptsize
    \centering
    \caption{List of courses at Miami University to cover the Cybersecurity ACM guideline}
    \label{tab:course_list}
    \begin{tabular}{ll} \hline
    Course Number & Course Name                       \\ \hline
    CYB 134        & Introduction to Cybersecurity     \\
    CYB 234        & Sysadmin \& scripting             \\
    CYB 235        & Computer networks                 \\
    CYB 236        & Data Security                     \\
    CYB 331        & Software Security                 \\
    CYB 332        & Organizational and Human Security \\
    CYB 334        & Network security                  \\
    CYB 335        & Defensive Security                \\
    CYB 435        & Offensive Security               \\ \hline
    \end{tabular}
\end{table}

\begin{table*}[h!]
\centering
\caption{List of recommended Cybersecurity topics and knowledge mapped to Miami's curriculum and related Learning Outcome Bloom's Taxonomy Level}
\label{tab:ku_tab}
\scriptsize
\begin{tabular}{|l|l|ll|}
\hline
\multirow{2}{*}{\bf Knowledge Areas}          & \multirow{2}{*}{\bf Knowledge Units}                                    & \multicolumn{2}{l|}{\bf Course coverage}           \\ \cline{3-4} 
                                          &                                                                     & \multicolumn{1}{l|}{\bf Course Label} & \bf LO   level \\ \hline
\multirow{10}{*}{Data Security}           & Cryptography                                                        & \multicolumn{1}{l|}{CYB 236}      & Apply      \\ \cline{2-4} 
                                          & Digital Forensics                                                   & \multicolumn{1}{l|}{CYB 236}      & Remember   \\ \cline{2-4} 
                                          & \multirow{2}{*}{Data Integrity and Authentication}                  & \multicolumn{1}{l|}{CYB 334}      & Apply      \\ \cline{3-4} 
                                          &                                                                     & \multicolumn{1}{l|}{CYB 236}      & Analyze    \\ \cline{2-4} 
                                          & \multirow{2}{*}{Access Control}                                     & \multicolumn{1}{l|}{CYB 334}      & Apply      \\ \cline{3-4} 
                                          &                                                                     & \multicolumn{1}{l|}{CYB 236}      & Understand \\ \cline{2-4} 
                                          & Secure Communication Protocols                                      & \multicolumn{1}{l|}{CYB 334}      & Analyze    \\ \cline{2-4} 
                                          & Cryptanalysis                                                       & \multicolumn{1}{l|}{CYB 236}      & Understand \\ \cline{2-4} 
                                          & Data Privacy                                                        & \multicolumn{1}{l|}{CYB 236}      & Understand \\ \cline{2-4} 
                                          & Information Storage Security                                        & \multicolumn{1}{l|}{CYB 236}      & Understand \\ \hline
\multirow{7}{*}{Software Security}        & Fundamental   Principles                                            & \multicolumn{1}{l|}{CYB 331}      & Understand \\ \cline{2-4} 
                                          & Design                                                              & \multicolumn{1}{l|}{CYB 331}      & Understand \\ \cline{2-4} 
                                          & Implementation                                                      & \multicolumn{1}{l|}{CYB 331}      & Apply      \\ \cline{2-4} 
                                          & Analysis and Testing                                                & \multicolumn{1}{l|}{CYB 331}      & Apply      \\ \cline{2-4} 
                                          & Deployment and Maintenance                                          & \multicolumn{1}{l|}{CYB 331}      & Understand \\ \cline{2-4} 
                                          & Documentation                                                       & \multicolumn{1}{l|}{CYB 331}      & Understand \\ \cline{2-4} 
                                          & Ethics                                                              & \multicolumn{1}{l|}{CYB 331}      & Understand \\ \hline
\multirow{5}{*}{Component Security}       & Component   Design                                                  & \multicolumn{1}{l|}{CYB 331}      & Understand \\ \cline{2-4} 
                                          & Component Fabrication                                               & \multicolumn{1}{l|}{CYB 331}      & Apply      \\ \cline{2-4} 
                                          & Component Procurement                                               & \multicolumn{1}{l|}{CYB 331}      & Remember   \\ \cline{2-4} 
                                          & Component Testing                                                   & \multicolumn{1}{l|}{CYB 331}      & Apply      \\ \cline{2-4} 
                                          & Component Reverse Engineering                                       & \multicolumn{1}{l|}{CYB 331}      & Remember   \\ \hline
\multirow{8}{*}{Connection Security}      & Physical   Media                                                    & \multicolumn{1}{l|}{CYB 235}      & Remember   \\ \cline{2-4} 
                                          & Physical Interfaces and Connectors                                  & \multicolumn{1}{l|}{CYB 235}      & Remember   \\ \cline{2-4} 
                                          & Hardware Architecture                                               & \multicolumn{1}{l|}{CYB 235}      & Remember   \\ \cline{2-4} 
                                          & Distributed Systems Architecture                                    & \multicolumn{1}{l|}{CYB 235}      & Remember   \\ \cline{2-4} 
                                          & Network Architecture                                                & \multicolumn{1}{l|}{CYB 235}      & Analyze    \\ \cline{2-4} 
                                          & Network Implementations                                             & \multicolumn{1}{l|}{CYB 235}      & Apply      \\ \cline{2-4} 
                                          & Network Services                                                    & \multicolumn{1}{l|}{CYB 235}      & Remember   \\ \cline{2-4} 
                                          & Network Defense                                                     & \multicolumn{1}{l|}{CYB 334}      & Apply      \\ \hline
\multirow{8}{*}{System Security}          & System   Thinking                                                   & \multicolumn{1}{l|}{CYB 335}      & Remember   \\ \cline{2-4} 
                                          & System Management                                                   & \multicolumn{1}{l|}{CYB 335}      & Understand \\ \cline{2-4} 
                                          & \multirow{2}{*}{System Access}                                      & \multicolumn{1}{l|}{CYB 334}      & Apply      \\ \cline{3-4} 
                                          &                                                                     & \multicolumn{1}{l|}{CYB 234}      & Understand \\ \cline{2-4} 
                                          & System Control                                                      & \multicolumn{1}{l|}{CYB 234}      & Understand \\ \cline{2-4} 
                                          & System Retirement                                                   & \multicolumn{1}{l|}{CYB 335}      & Remember   \\ \cline{2-4} 
                                          & System Testing                                                      & \multicolumn{1}{l|}{CYB 435}      & Apply      \\ \cline{2-4} 
                                          & Example System Architectures                                        & \multicolumn{1}{l|}{CYB 435}      & Understand \\ \hline
\multirow{7}{*}{Human Security}           & Identity   Management                                               & \multicolumn{1}{l|}{CYB 332}      & Understand \\ \cline{2-4} 
                                          & Social Engineering                                                  & \multicolumn{1}{l|}{CYB 332}      & Understand \\ \cline{2-4} 
                                          & Personal Compliance with Cybersecurity   Rules/Policy/Ethical Norms & \multicolumn{1}{l|}{CYB 332}      & Remember   \\ \cline{2-4} 
                                          & Awareness and Understanding                                         & \multicolumn{1}{l|}{CYB 332}      & Remember   \\ \cline{2-4} 
                                          & Social and Behavioral Privacy                                       & \multicolumn{1}{l|}{CYB 332}      & Understand \\ \cline{2-4} 
                                          & Personal Data Privacy and Security                                  & \multicolumn{1}{l|}{CYB 332}      & Understand \\ \cline{2-4} 
                                          & Usable Security and Privacy                                         & \multicolumn{1}{l|}{CYB 332}      & Understand \\ \hline
\multirow{13}{*}{Organizational Security} & \multirow{2}{*}{Risk Management}                                    & \multicolumn{1}{l|}{CYB 335}      & Understand \\ \cline{3-4} 
                                          &                                                                     & \multicolumn{1}{l|}{CYB 332}      & Understand \\ \cline{2-4} 
                                          & \multirow{2}{*}{Security Governance \& Policy}                      & \multicolumn{1}{l|}{CYB 435}      & Understand \\ \cline{3-4} 
                                          &                                                                     & \multicolumn{1}{l|}{CYB 332}      & Remember   \\ \cline{2-4} 
                                          & Analytical Tools                                                    & \multicolumn{1}{l|}{CYB 335}      & Understand \\ \cline{2-4} 
                                          & \multirow{2}{*}{Systems Administration}                             & \multicolumn{1}{l|}{CYB 335}      & Apply      \\ \cline{3-4} 
                                          &                                                                     & \multicolumn{1}{l|}{CYB 234}      & Apply      \\ \cline{2-4} 
                                          & \multirow{2}{*}{Cybersecurity Planning}                             & \multicolumn{1}{l|}{CYB 335}      & Understand \\ \cline{3-4} 
                                          &                                                                     & \multicolumn{1}{l|}{CYB 332}      & Remember   \\ \cline{2-4} 
                                          & Business Continuity, Disaster Recovery, and   Incident Management   & \multicolumn{1}{l|}{CYB 335}      & Remember   \\ \cline{2-4} 
                                          & Security Program Management                                         & \multicolumn{1}{l|}{CYB 335}      & Understand \\ \cline{2-4} 
                                          & Personnel Security                                                  & \multicolumn{1}{l|}{CYB 335}      & Remember   \\ \cline{2-4} 
                                          & Security Operations                                                 & \multicolumn{1}{l|}{CYB 335}      & Understand \\ \hline
\multirow{8}{*}{Societal Security}        & \multirow{2}{*}{Cybercrime}                                         & \multicolumn{1}{l|}{CYB 435}      & Understand \\ \cline{3-4} 
                                          &                                                                     & \multicolumn{1}{l|}{CYB 134}      & Remember   \\ \cline{2-4} 
                                          & Cyber Law                                                           & \multicolumn{1}{l|}{CYB 134}      & Remember   \\ \cline{2-4} 
                                          & \multirow{2}{*}{Cyber Ethics}                                       & \multicolumn{1}{l|}{CYB 435}      & Understand \\ \cline{3-4} 
                                          &                                                                     & \multicolumn{1}{l|}{CYB 134}      & Remember   \\ \cline{2-4} 
                                          & Cyber Policy                                                        & \multicolumn{1}{l|}{CYB 134}      & Remember   \\ \cline{2-4} 
                                          & \multirow{2}{*}{Privacy}                                            & \multicolumn{1}{l|}{CYB 234}      & Apply      \\ \cline{3-4} 
                                          &                                                                     & \multicolumn{1}{l|}{CYB 134}      & Remember   \\ \hline
\end{tabular}
\end{table*}

Table~\ref{tab:course_list} shows our Miami University's cybersecurity courses \cite{course_list} and the course name.  The topics covered for ACM's cybersecurity curriculum \cite{acm_guideline} are presented in Table~\ref{tab:ku_tab}.   In Columns 1 and 2, we show the knowledge area (a larger grouping) and knowledge units, where there are several knowledge units per area.  Column 3 shows which course covers a particular knowledge unit, and Column 4 lists the Bloom taxonomy level for the learning outcome for each knowledge unit.  We do not list the exact LO for each knowledge unit, and instead, we write out the LO when we talk about how the verbiage is potentially increased for a particular example.

The interesting information from this table is a summary view of each of all the knowledge units and noting how ``low-order'' the majority of knowledge units are with respect to Bloom's Taxonomy.  For the first three levels of ``1 - Remember'', ``2 - Understand'', and ``3 - Apply'' in Table~\ref{tab:ku_tab}, there are 22, 28, and 14 instances of these, totaling 65 of the 66 listed course LOs.  The highest Bloom's level for our curriculum includes 1 instance of ``4 - Analyze''. 

The main reason for the above result is the number of LOs listed for covering the cybersecurity curriculum over nine courses.  The reality is there is a lot of ideas to learn about in the space of cybersecurity as a system, and its complexities (beyond hardware and software) make technological and psychological knowledge and skills a tremendous burden on this type of education.  As we've alluded to in this paper, the question is, will tools such as ChatGPT allow us to push our learners and teachers into an accelerated space?

\section{GPT Examples where the Tool may allow us to Raise and Accelerate Learning Outcomes}

As noted in the last section's analysis, the demands of remembering, understanding, and applying new knowledge and skills are overwhelming for the cybersecurity curriculum.   This section provides a few examples of how an LO can be linked with ChatGPT to accelerate and potentially raise the skill and knowledge to a higher Bloom's taxonomy level.  As stated earlier, the goal is not to simply move up in terms of Bloom's Taxonomy levels as this is not a feasible approach to learning and becoming an effective practitioner.  Instead, our goal is to provide examples of how ChatGPT can be used to extend what the learner achieves in terms of the knowledge unit.  In each example, we will show the knowledge unit in the title, show the prompt and response and then discuss how this allows the learner to push their knowledge further.

\subsection{Data Security - Access Control}

The original LO for this knowledge unit is ``Identify and explain the importance of basic system administration tasks'' in the course CYB 234 \footnote{https://www.miamioh.edu/cec/academics/departments/cse/academics/course-descriptions/cyb234/index.html}.

\begin{figure}[h]
    \centering
    \begin{adjustbox}{minipage=[t]{1\linewidth},fbox}
        \textbf{Prompt:} What is the Linux command to add execute permission to just group users
        
        \textbf{Response:} The command to add execute permission to just group users in Linux is 
            \begin{lstlisting}[style=customc,breaklines=true]
chmod g+x <filename>
            \end{lstlisting} 
Here, ``chmod'' is the command used to change file permissions, ``g'' stands for a group, and ``x'' adds execute permission to the file for the group. Replace $<$filename$>$ with the name of the file for which you want to add execute permission to the group.

    \end{adjustbox}
    \caption{Prompt and response for using Linux file permissions}
    \label{fig:chatgpt-example-1}
\end{figure}

Figure~\ref{fig:chatgpt-example-1} shows an example of how to move from \textbf{Understand} to \textbf{Apply} in data security access control.  Instead of just describing how the system permissions allow users to perform certain actions in Linux, we now have details of what is the exact Linux syntax for the command line and can now test it in the tool.

With ChatGPT acting as a tutorial agent, we can imagine a student being able to test out on a Linux system how the command works and observe what the system does to manage access control for a shared system.

\subsection{Software Security Analysis and Testing}

The original LO for this knowledge unit is ``Describe and apply static and dynamic software testing methods'' in CYB 331 - Software Security.

\begin{figure}[h]
    \centering
    \begin{adjustbox}{minipage=[t]{1\linewidth},fbox}
            \textbf{Prompt:} Does the following program have a memory leak problem?
            \begin{lstlisting}[style=customc,breaklines=true]
void f() {
   int *ptr = (int *) malloc(sizeof(int));
   /* Do some work */
   return; /* Return without freeing ptr*/
}
            \end{lstlisting}
           \textbf{Response:} Yes, the program has a memory leak problem. The ``malloc()'' function allocates memory dynamically, but the ``ptr'' pointer is never freed, which means that the memory remains allocated even after the function returns. This can lead to memory leaks and other issues if the program continues to allocate memory without freeing it. To fix this problem, the memory allocated by malloc() should be freed before the function returns using the free() function.  Here's an example of how to modify the code to fix the memory leak:
           \begin{lstlisting}[style=customc,breaklines=true]
void f() {
    int *ptr = (int ) malloc(sizeof(int));
    /* Do some work */
    free(ptr); /* Free the memory before returning */
    return;
}
            \end{lstlisting}
    \end{adjustbox}
    \caption{Prompt and response for analyzing some C-code}
    \label{fig:chatgpt-example-2}
\end{figure}

Figure~\ref{fig:chatgpt-example-2} shows how we can use ChatGPT to accelerate a student's understanding of software security analysis from \textbf{Apply} to \textbf{Analysis}. 
 With the ChatGPT, the user can prompt the tool to analyze their code and make it more secure.

We imagine providing code to the student that shows a number of cases such as memory leaks and buffer overflows that ChatGPT can help them analyze and understand.  Next, given a larger code base, either their own or a manufactured case, could the learner use the tool to analyze the code, and then help the student create a safer version of it.  

\subsection{Component reverse engineering}

The original LO for this knowledge unit is ``Explain how failures in fundamental security design principles can lead to system vulnerabilities that can be exploited as part of an offensive cyber operation'' in CYB 331 - Software Security.

\begin{figure}[h]
    \centering
    \begin{adjustbox}{minipage=[t]{1\linewidth},fbox}
            \textbf{Prompt:} Reverse engineer the following code to C code: 
            \begin{lstlisting}[style=customc,breaklines=true]
fact(int):
    movl $1, %eax 
    cmpl $1 %edi 
    jle .L1 
.L2: movl %edi, %edx 
    subl $1, %edi 
    imull %edx, %eax 
    cmpl $1 %edi 
    jne .L2 
.L1: ret"
            \end{lstlisting}
           \textbf{Response:} Sure, here's the equivalent C code for the given x86 assembly code:
           \begin{lstlisting}[style=customc,breaklines=true]
int fact(int x) {
    int result = 1;
    if (x <= 1) {
        return result;
    } else {
        do {
            result *= x;
            x--;
        } while (x > 1);
        return result;
    }
}   
            \end{lstlisting}
    The fact function takes an integer argument x and returns the factorial of x. Here's how the x86 code translates to C code:

\begin{itemize}
    \item  movl \$1, \%eax sets the eax register to 1, which initializes the result variable to 1.
    \item cmpl \$1, \%edi compares the value of x with 1.
    \item jle .L1 jumps to the .L1 label if x is less than or equal to 1. This is the base case of the factorial function.
    \item .L2: is a label that marks the start of the loop.
    \item movl \%edi, \%edx moves the value of x into the edx register, which is used to calculate the next iteration of the result.
    \item subl \$1, \%edi decrements x by 1.
    \item imull \%edx, \%eax multiplies the current value of result (stored in the eax register) with the current value of x (stored in the edx register), and stores the result back into eax.
    \item cmpl \$1, \%edi compares the new value of x with 1.
    \item jne .L2 jumps back to .L2 label if x is not equal to 1, which means that we need to continue the loop to calculate the next iteration of the result.
    \item .L1: is the label that we jump to if x is less than or equal to 1. This is the base case of the factorial function.
    \item ret returns the result of the function to the caller.
\end{itemize}
    \end{adjustbox}
    \caption{Prompt and response for analyzing some C-code}
    \label{fig:chatgpt-example-3}
\end{figure}

Figure~\ref{fig:chatgpt-example-3} shows what ChatGPT provides when prompted to reverse engineer some assembly code.  This allows students to go beyond remembering what reverse engineering is and into the space of understanding what potentially is possible with using reverse engineering.  

In this space, the tool provides rudimentary capabilities in reverse engineering.  The question is, can students reverse engineer other more complex pieces of code, and similarly, obfuscate their own code examples such that they can fool this simple tool?

\subsection{Connection Security - Distributed System Architecture}

The original LO for this knowledge unit is ``Describe application layer security protocols'' in CYB 334\footnote{https://www.miamioh.edu/cec/academics/departments/cse/academics/course-descriptions/cyb334/index.html}.

\begin{figure}[h]
    \centering
    \begin{adjustbox}{minipage=[t]{1\linewidth},fbox}
           \textbf{Prompt:} Write a command that will find all computers in a network running MySQL
            
           \textbf{Response:} Assuming you have nmap installed on your system, you can run the following command:
           \begin{lstlisting}[style=customc,breaklines=true]
sudo nmap -p 3306 --open -sV <network-address> 
            \end{lstlisting}
    Replace <network-address> with the IP address or the network address of the network that you want to scan for devices running MySQL. For example, if the network address is 192.168.0.0/24, you would run:

           \begin{lstlisting}[style=customc,breaklines=true]
sudo nmap -p 3306 --open -sV 192.168.0.0/24
            \end{lstlisting}
          This command will scan the IP addresses in the specified network range and check if port 3306 (default MySQL port) is open on each of them. It will also attempt to determine the version of MySQL running on each device.

            Note that you may need to modify the IP address range according to your network configuration. Additionally, the sudo command is used to run nmap with elevated privileges, which may be necessary depending on your system configuration. 
        \end{adjustbox}
        
    \caption{Prompt and response for looking at a network to find out what servers are accepting}
    \label{fig:chatgpt-example-4}
\end{figure}

\begin{figure}[h]
    \centering
    \begin{adjustbox}{minipage=[t]{1\linewidth},fbox}
           \textbf{Prompt:} What linux commands can help a sys admin to evaluate their network security?
            
           \textbf{Response:} Here are some Linux commands that can help a system administrator evaluate their network security:
           \begin{enumerate}
               \item  nmap: nmap is a powerful network scanner that can be used to identify open ports, running services, and detect vulnerabilities in a network. It can be used to scan specific hosts or entire networks.
                \item netstat: netstat is a command-line tool used for monitoring network connections. It can show open network sockets, their associated processes, and network interface statistics.
                \item tcpdump: tcpdump is a packet analyzer that can capture and display network traffic in real-time. It can be used to inspect packets for potential security issues and can also be used to identify suspicious network activity.
                \item iptables: iptables is a command-line utility used for configuring firewall rules in Linux. It can be used to create rules that filter or block network traffic based on specific criteria, such as IP address or port number.
                \item fail2ban: fail2ban is a log analysis and intrusion prevention tool. It can be used to monitor log files for suspicious activity and automatically block IP addresses that show signs of malicious activity.
                \item chkrootkit: chkrootkit is a tool that can be used to scan for rootkits on a Linux system. It checks for common signs of a rootkit infection, such as hidden files and processes.
                \item lynis: lynis is a security auditing tool that can be used to scan Linux systems for security vulnerabilities. It can perform a variety of checks, including system and file permissions, firewall configuration, and malware scanning.
           \end{enumerate}
            These commands can be combined with other security tools and practices to help ensure the security of a Linux network. However, it's important to note that evaluating network security is a complex and ongoing process that requires a comprehensive approach. 
        \end{adjustbox}
        
    \caption{Prompt and response for a query on network security tools in Linux}
    \label{fig:chatgpt-example-5}
\end{figure}

Figure~\ref{fig:chatgpt-example-4} shows what ChatGPT can describe in terms of using a Linux system to evaluate the state of the network.  Using a simple ChatGPT query based on Figure~\ref{fig:chatgpt-example-5} allows a learner even to understand what are the possible system administrative tools for network security.  Again, using ChatGPT as a personalized tutor allows the learner to get direct feedback on how to proceed in the space.  

\section{Conclusion}

In this work, we looked at our existing cybersecurity curriculum as a case study of how well AI-driven natural language processing tools  such as ChatGPT change the LOs of a curriculum.  To do this we used the lense of LOs and their respective Bloom's taxonomy verbiage and mapped out our cybersecurity curriculum.  We noticed that there are a large number of LOs in Cybersecurity and most of those LOs sit in the ``low-order'' thinking skills.  Based on this, we hypothesize that ChatGPT will allow us to accelerate some of our existing LOs, given the tool's capabilities.  To show this we provide five examples of how an existing LO can be accelerated with ChatGPT, given the capabilities of the tool.

From this exercise, we have learned two things in particular that we believe we will need to be further examined by all educators.  First, our experiences with ChatGPT suggest that the tool can provide a powerful means to allow learners to generate pieces of their work quickly.  However, learners will need to be taught how to create prompts, and how to validate the responses and designs provided by the tool, and just like using the browser to find things, the learner needs to be aware of the vocabulary used to get acceptable responses.  Second, we will need to consider how to teach concepts that need to be experienced from ``first-principle'' learning approaches and learn how to motivate students to perform some rudimentary exercises that ``the tool'' can easily do for me.  We will argue, that some basic experiments need to be done that provide us with evidence of why certain concepts still need to be done to help us prove to our students the importance of early learning exercises.

The experiment we suggest going forward is to use the cybersecurity concept inventory for a control class that is taught various cyber concepts in a traditional method and compares a group that is taught similar ideas but is also prompted to use ChatGPT to explore the ideas more deeply.  This depth could be facilitated by designing assignments that ask for more detailed challenges within the cybersecurity space.  

\bibliographystyle{IEEEtran}
\bibliography{joined.bib}

\end{document}